\RequirePackage[2020-02-02]{latexrelease}
\documentclass[aps,prc,twocolumn,letterpaper,10pt,twoside,superscriptaddress,tightenlines,nofootinbib,showpacs]{revtex4}
\usepackage{amsmath,amsfonts,amssymb,latexsym}
\usepackage{graphicx,color}
\usepackage[sort&compress]{natbib}

\begin{document}

%%%   New Definitions
\newcommand{\eg}{{\it e.g.}}
\newcommand{\cf}{{\it cf.}}
\newcommand{\etal}{{\it et. al.}}
\newcommand{\ie}{{\it i.e.}}
\newcommand{\be}{\begin{equation}}
\newcommand{\ee}{\end{equation}}
\newcommand{\bea}{\begin{eqnarray}}
\newcommand{\eea}{\end{eqnarray}}
\newcommand{\bef}{\begin{figure}}
\newcommand{\eef}{\end{figure}}
\newcommand{\bce}{\begin{center}}
\newcommand{\ece}{\end{center}}
\newcommand{\red}[1]{\textcolor{red}{#1}}

\newcommand{\dd}{\text{d}}
\newcommand{\ii}{\text{i}}
\newcommand{\lsim}{\lesssim}
\newcommand{\gsim}{\gtrsim}
\newcommand{\RAA}{R_{\rm AA}}
\newcommand{\sig}{\sigma_{\rm el}}
\newcommand{\Cem}{C_{\rm em}}
\newcommand{\rhoem}{\rho_{\rm em}}
\newcommand{\Vfb}{V_{\rm FB}}

\title{Electric Conductivity of QCD Matter and Dilepton Spectra in Heavy-Ion Collisions}

%\author{Joseph~Atchison}
%\address{Department of Physics, Abilene Christian University, Abilene, TX, U.S.A.}
%\address{Cyclotron Institute and Department of Physics and
%Astronomy, Texas A\&M University, College Station, Texas 77843-3366, U.S.A.}
\author{Ralf~Rapp}
\address{Cyclotron Institute and Department of Physics and
Astronomy, Texas A\&M University, College Station, Texas 77843-3366, U.S.A.}

\date{\today}

\begin{abstract}
The electric conductivity, $\sig$, is a fundamental transport coefficient of QCD matter that can be related to the zero-energy limit of the electromagnetic (EM)
spectral function at vanishing 3-momentum in the medium. The EM spectral function is also the central quantity to describe the thermal emission rates and
pertinent spectra of photon and dilepton radiation in heavy-ion collisions. Employing a model for dilepton rates that combines hadronic many-body theory with nonperturbative QGP emission constrained by lattice-QCD which describes existing dilepton measurements in heavy-ion collisions, we investigate 
the sensitivity of  low-mass dilepton spectra in Pb-Pb collisions at the LHC to $\sig$. In particular, we disentangle the contributions from QGP and hadronic 
emission, and identify signatures that can help to extract $\sig$ from high-precision experimental data expected to be attainable with future detector 
systems at the LHC. 
\end{abstract}

%\pacs{25.75.-q  25.75.Dw  25.75.Nq}

\maketitle

%%%%%%%%%%%%%%%%%%%%%%%%%%%%%%%
\section{Introduction}
%%%%%%%%%%%%%%%%%%%%%%%%%%%%%%%
The hot QCD medium as produced in experiments using ultra-relativistic heavy-ion collisions (URHICs) exhibits remarkable transport properties. Most 
of the information on those to date has been deduced from phenomenological extractions of the shear viscosity, $\eta$, characterizing the transport 
of energy-momentum, and the charm-diffusion coefficient, ${\cal D}_s$, characterizing the transport of the heavy-flavor (HF) quantum number, 
within the expanding medium. Utilizing relativistic hydrodynamic~\cite{Heinz:2013th,Niemi:2015qia} and Brownian motion  
approaches~\cite{Rapp:2018qla,Dong:2019byy} to simulate the momentum spectra of light hadrons and of HF particles, respectively, 
the dimensionless ratios, $4\pi\eta/s$ ($s$:  entropy density) and $2\pi T{\cal D}_s$ ($T$: temperature), have been estimated to be as small as a 
factor 2 within lower bounds conjectured for the strong-coupling limit of quantum field theory~\cite{Policastro:2002se}.  Much less is known about 
the electric conductivity, $\sig$, where only recently phenomenological analyses using charge balance function have been conducted, while theoretical
calculations in both hadronic and quark-gluon 
matter~\cite{Fernandez-Fraile:2009eug,Marty:2013ita,Lee:2014pwa,Greif:2014oia,Greif:2016skc,Ghosh:2016yvt}, including lattice-QCD (lQCD) computations~\cite{Aarts:2014nba,Brandt:2015aqk,Ding:2016hua}, give rather widely varying results, by over a factor of $\sim$5.

In thermal field theory transport coefficients can be related to the low-energy limit of the zero-momentum spectral function of the pertinent current
of the conserved quantity (\eg, energy-momentum, flavor or charge). For the electric conductivity this corresponds to the (spatial components of the) electromagnetic (EM) spectral function, $\rhoem$. The latter, in turn, is the only spectral function whose medium modifications can be directly measured in 
nuclear-collision experiments, through the radiation of dileptons which, due to their long mean-free path, can penetrate the strongly interacting system
essentially unscathed. Over the past $\sim$25 years, a robust understanding of the observed low-mass excess of dilepton spectra in heavy-ion collisions 
over a large range of energies has been achieved~\cite{Rapp:2009yu,Rapp:2016xzw}. A main conclusion is that the $\rho$-meson, which provides 
the dominant contribution to hadronic dilepton production, strongly broadens in hadronic matter, with little mass shift, to the extent that the resonance 
peak ``melts" into a quark-antiquark continuum at temperatures near 170\,MeV, thereby indicating a transition to partonic degrees of freedom.
The EM spectral functions have also been extrapolated to the light cone, \ie, to vanishing invariant mass $M$=0, to extract thermal photon rates which 
have been utilized in the interpretation of experimental data for direct-photon production~\cite{Turbide:2003si,vanHees:2014ida,Paquet:2015lta}.

The purpose of the present paper is to carry the in-medium  EM spectral function employed in heavy-ion phenomenology to the timelike low-energy limit 
and (a) extract the pertinent conductivity, and (b) investigate its signatures in dilepton measurements at very low mass and momentum. In this way, we 
elaborate connections between the microscopic processes that drive the low-mass enhancement of dilepton production in experiment, typically observed
in the mass region of a few hundred MeV, and charge transport in the medium.

The remainder of this article is organized as follows.  In Sec.~\ref{sec_sf} we recall the basic ingredients to the EM spectral function consisting of 
in-medium vector-meson spectral functions in hadronic matter and a lQCD constrained rate in the quark-gluon plasma (QGP). 
In Sec.~\ref{sec_cond} we study the conductivity peak near zero energy, in particular its collisional broadening and the resulting reduction of the 
conductivity as the zero-energy limit, as well as its manifestation in thermal photon emission rates. 
In Sec.~\ref{sec_dl-spec} we investigate the sensitivity of thermal dilepton spectra in Pb-Pb collisions at the LHC at very low mass ($M\lsim0.2$~GeV) 
to different scenarios of the conductivity in the underlying emission rates; specifically, we show how different acceptance cuts on the single-electron as well as 
electron pair momenta affect the experimental observables. We summarize and conclude in Sec.~\ref{sec_concl}.

%%%%%%%%%%%%%%%%%%%%%%%%%%%%%%%
\section{EM Spectral Functions of QCD Matter}
\label{sec_sf}
%%%%%%%%%%%%%%%%%%%%%%%%%%%%%%%
The thermal rate for dilepton radiation off strongly interacting matter, per unit four-momentum and four-volume, can be expressed as
\be
\frac{dN_{l^+l^-}}{d^4qd^4x}=  \frac{\alpha^2 L(M)}{2\pi^3M^2} \rhoem(M,q;\mu_B,T)
\ee  
where the EM spectral function, $\rhoem$,  is related to the imaginary part of the EM current-current correlation function as 
$\rhoem \equiv - 2 g_{\mu\nu}{\rm Im}\Pi_{\rm em}^{\mu\nu}/3 \equiv -2 ({\rm Im} \Pi_L + 2 {\rm Im} \Pi_T)/3$. Here, $M^2=q_0^2-q^2$  and $q$ 
denote the invariant mass squared and three-momentum of the virtual photon,  and $L(M)$ a final-state lepton phase space factor (either for di-electrons or 
di-muons).

For the emission rates from hadronic matter, we employ the the in-medium vector meson spectral functions developed in Refs.~\cite{Urban:1999im,Rapp:1999us,Rapp:2000pe}. We focus on the isovector ($\rho$-meson) channel, which gives the dominant contribution (the contributions of in-medium 
$\omega$ mesons in the isoscalar channel are included in the calculations of dilepton spectra discussed in Sec.~\ref{sec_dl-spec}). The starting point is 
an effective, gauge-invariant lagrangian for $\pi\rho$ interactions in vacuum, whose parameters (bare $\rho$ mass, $m_\rho^{(0)}$,  $\rho\pi\pi$ coupling 
constant, $g_\rho$ and a pertinent vertex formfactor cutoff, $\Lambda_\rho$) are fitted to reproduce experimental data for $P$-wave $\pi\pi$ scattering 
phase shifts, the pion EM formfactor and the EM spectral function measured in $e^+e^-$ annihilation into hadrons as well as in hadronic $\tau$ decays,
utilizing the vector dominance model, 
\begin{equation}
\rhoem(M) = - 2\frac{(m_\rho^{(0)})^4}{g_\rho^2} {\rm Im} D_\rho(M)  \ .
\end{equation} 
Medium effects are calculated based on interactions of the $\rho$ meson and its pion cloud with the surrounding hadronic medium using quantum 
many-body theory employing effective lagrangians whose parameters are constrained by empirical decay branchings and scattering data, \eg,  
resonance decays of mesons (\eg, $a_1\to\rho\pi$) and baryons  ($B^*\to N\rho$) and their radiative decays, photo-absorption on the nucleon and 
nuclei, and $\pi N \to\rho N$ scattering. 
%in particular, it also provides a good description of the   

For the emission rate from the QGP we employ the construction detailed in Ref.~\cite{Rapp:2013nxa}. Following the lQCD analysis in 
Ref.~\cite{Ding:2010ga}, perturbative $q\bar q$ annihilation can been augmented by a low-energy part using a Breit-Wigner peak with 2 parameters 
adjusted to reproduce lQCD data of Euclidean correlation functions. This corresponds to the conductivity peak in the spectral function whose zero-energy 
limit determines the value of $\sig$. It turns out that this peak can also be simulated by an energy dependence that is given by the leading-order photon 
emission rate (which essentially corresponds to Bremsstrahlung and quark-gluon Compton  processes)~\cite{Rapp:2013nxa}. This implementation has 
the advantage that it generates a 3-momentum dependence of the dilepton rate. Furthermore, introducing a $K$-factor of 2 for this part, one can 
simultaneously account for higher-order effects in the emission rate and reproduce the low-energy conductivity peak rather well.

\begin{figure*}[!t]
\includegraphics[width=0.48\textwidth,angle=-0]{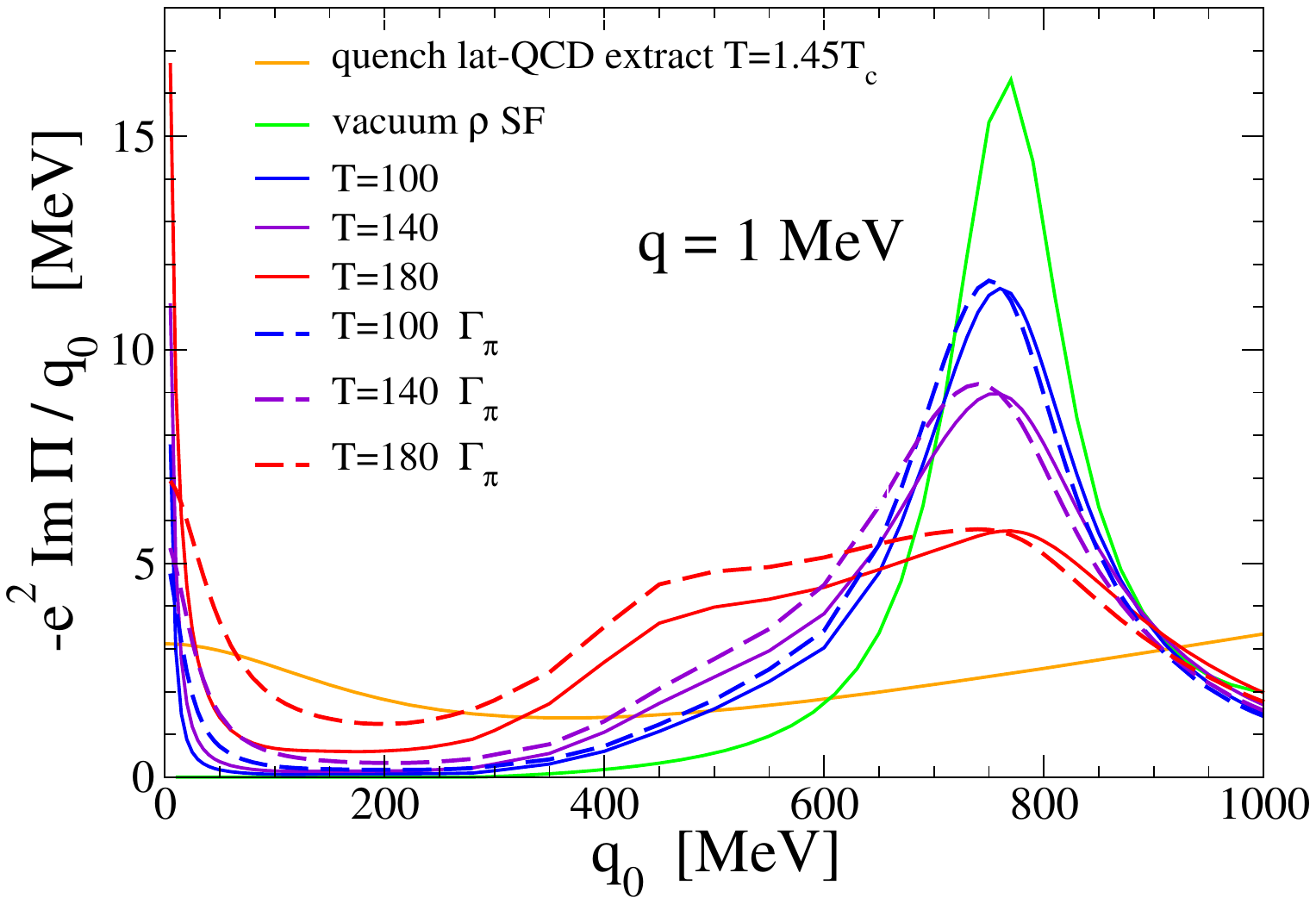}
\includegraphics[width=0.48\textwidth,angle=-0]{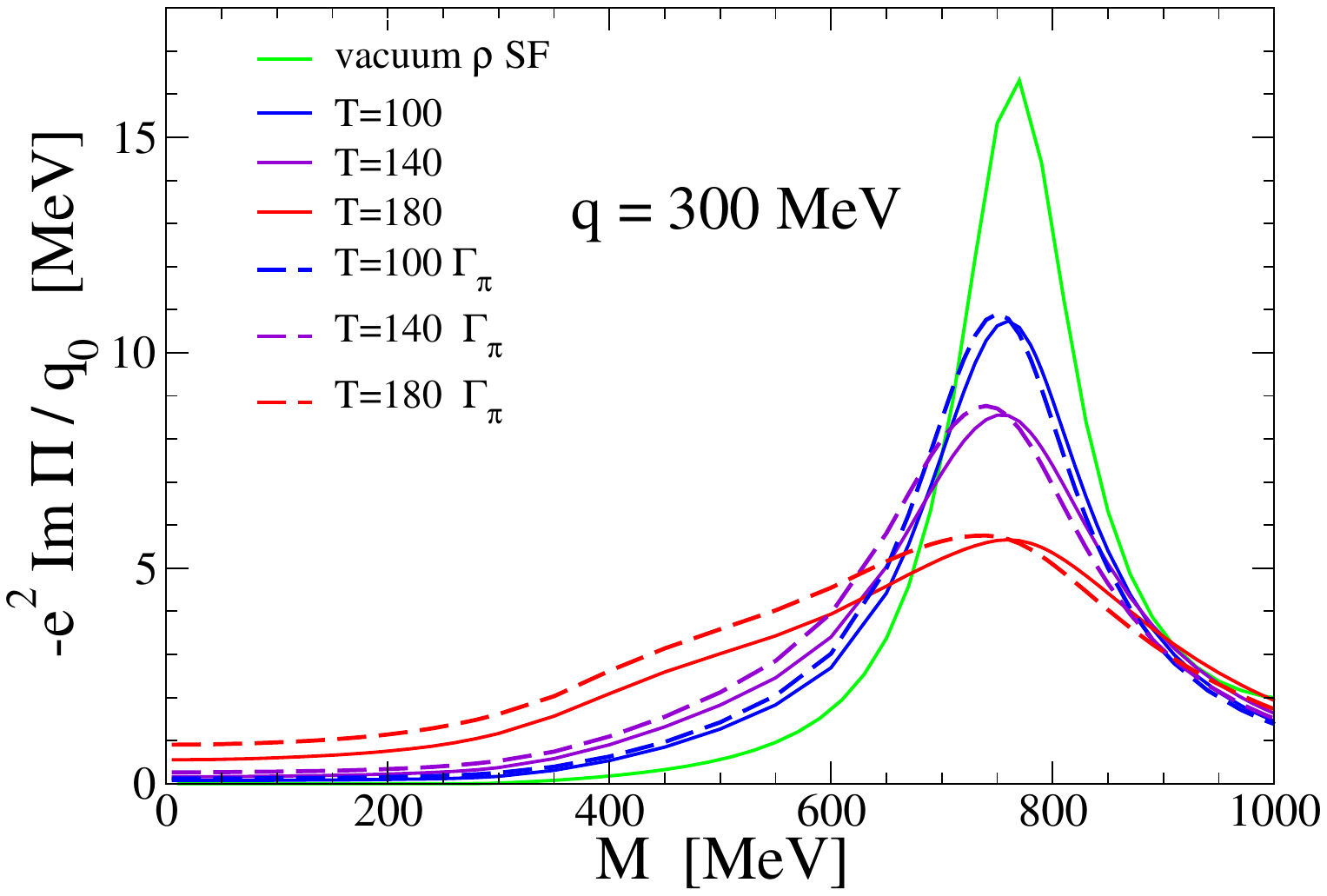}
\vspace{-0.2cm}
\caption{Electromagnetic spectral functions in hot hadronic matter obtained from in-medium $\rho$ propagators at temperatures and baryon chemical 
potentials along a thermodynamic trajectory at fixed entropy per baryon, $s/\varrho_B$=330, and chemical freezeout at ($T,\mu_B$)=(160,22)\,MeV 
(approximately reflecting the conditions in Au-Au(0.2\,TeV) collisions at RHIC). The solid lines are the results based on Ref.~\cite{Urban:1999im} while 
the dashed lines additionally include constant thermal-pion widths based on Ref.~\cite{Rapp:1995fv}.
Left panel: as function of energy at (near-) vanishing 3-momentum, also showing the QGP result at $T\simeq225$\,MeV (orange solid line); 
Right panel:  at 3 a finite 3-momentum of $q$=300\,MeV.  
Note that the spectral functions are divided by energy and multiplied by the electric-charge squared so that for vanishing momentum the zero-energy 
intercept with the abscissa correspond to the electric conductivity, $\sig$.}
\label{fig_rhoem-x}
\end{figure*}
Examples for the in-medium EM spectral functions are shown in Fig.~\ref{fig_rhoem-x}, normalized in a way that the zero-energy limit 
(at vanishing 3-momentum)  corresponds to the value of the electric conductivity, which will be investigated in more detail in the following section. 
The spectral functions in hadronic matter, essentially reflecting the in-medium $\rho$ spectral function, exhibit the well-known broadening of the 
$\rho$ peak as temperature and total baryon density increase, mostly driven by the effects from interactions with baryons and 
anti-baryons~\cite{Rapp:2000pe}. As an improvement over our earlier applications to EM observables, necessary to ensure a more reliable evaluation 
of the very-low-mass region, we have replaced the previously used 3-level scheme for the pion selfenergies from interactions with 
baryons by a full off-shell treatment  (the net effect turns out to be small at higher masses, as will be discussed more quantitatively in the context of 
dilepton spectra in Sec.~\ref{sec_dl-spec}).
In addition, we have augmented the medium modifications of the $\rho$'s pion cloud~\cite{Urban:1999im} by implementing (3-momentum averaged) pion 
widths generated by interactions with thermal pions in the heat bath, estimated from the studies in Refs.~\cite{Atchison:2022yxm,Rapp:1993ih}, 
amounting to $\sim$43, 26 and 17\,MeV at $T$=180, 140 and 100\,MeV, respectively. These effects have long been known to play a rather minor role in 
the medium modifications of the $\rho$ meson~\cite{Rapp:1995fv}, as confirmed here as well (compare the solid vs.~the dashed lines of the same 
color).\footnote{We note that the main medium effect from the surrounding thermal pions on the $\rho$'s pion cloud is caused by the Bose enhancement 
factors on the pion cloud~\cite{Rapp:1993ih}; this quantum-statistical effect has already been included in Refs.~\cite{Urban:1999im,Rapp:1999us}, 
corresponding to the solid lines in Fig.~\ref{fig_rhoem-x}.} 
However, they have been found to play a more important role at small energies and specifically for the conductivity~\cite{Atchison:2022yxm}.  
The QGP result (orange line in the  left panel of Fig.~\ref{fig_rhoem-x}) reflects the structureless $q\bar q$ continuum toward higher masses 
and a broad  conductivity peak at small energies.

\begin{figure}[!t]
\includegraphics[width=0.49\textwidth,angle=-0]{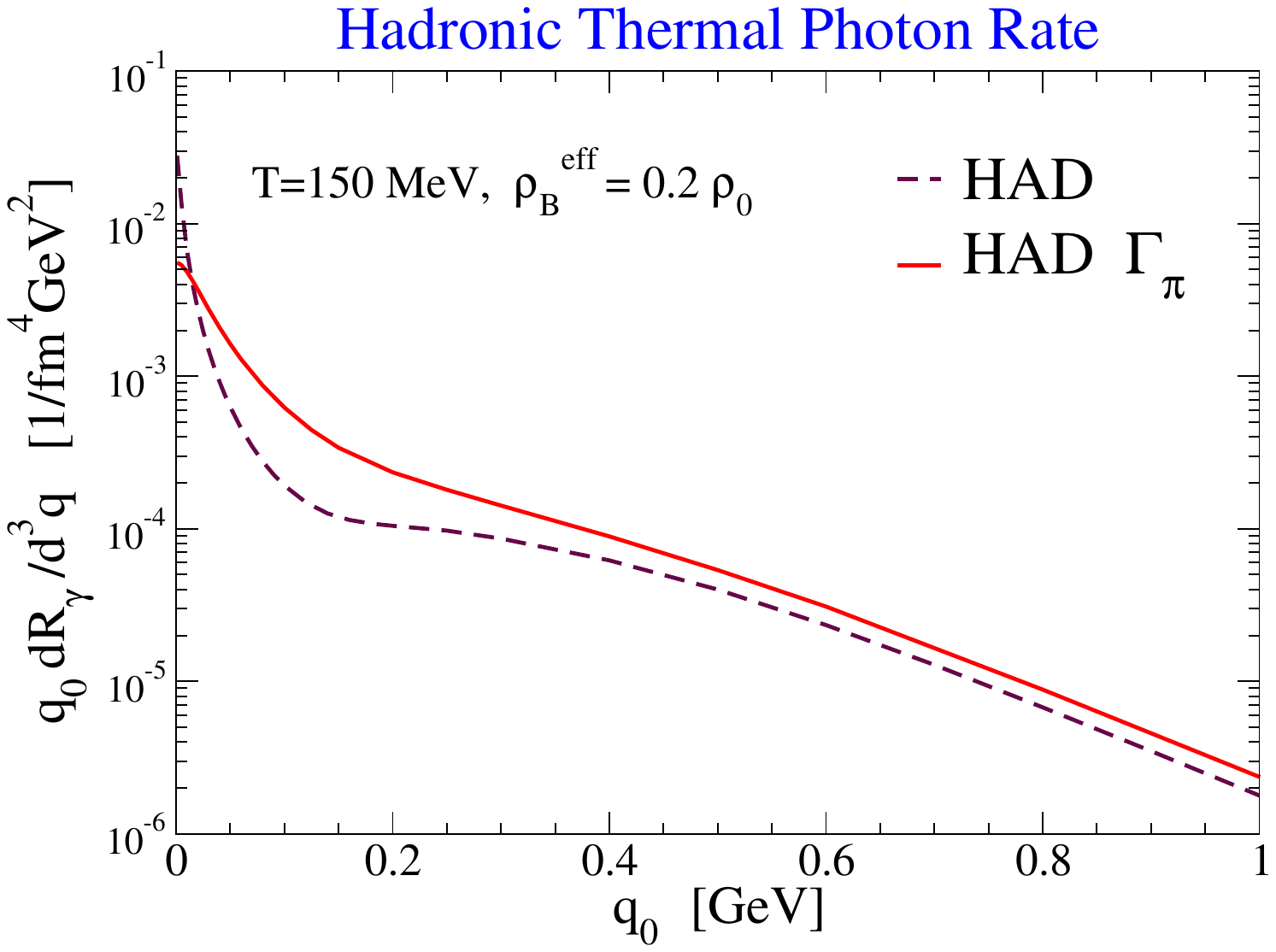}
\vspace{-0.2cm}
\caption{Thermal hadronic photon rate from the in-medium $\rho$ spectral function with (solid line) and without (dashed line) the contribution from $\pi\pi$ Bremsstrahlung approximated by a thermal-pion induced width in the pion propagators of the $\rho$-meson selfenergy.}
\label{fig_photon-rate}
\end{figure}
As a further illustration of the role of the pion widths generated by interactions with thermal pions (which are mostly mediated through the $\sigma$ and $\rho$ resonances),  we display in Fig.~\ref{fig_photon-rate} the thermal photon rate following from the in-medium $\rho$ spectral function in hadronic matter. The addition of thermal-pion widths has a significant impact on the low-energy rate for energies in the 100-300\,MeV region. This is a direct consequence of the broadening of the transport peak, which is accompanied by a quenching of the peak value at $q_0\to 0$ discernible only at extremely small energies of below 
$\sim$10\,MeV. This suggests that an experimentally measurable signature of a small conductivity is not a peak quenching but rather an enhancement 
from the peak broadening. The dominant processes underlying the thermal-pion widths are $\pi\pi$ Bremsstrahlung of photons, as can be inferred from 
pertinent cuts through the imaginary part of the $\rho$-meson's pion cloud selfenergy, with a thermal insertion on both pion lines. The pertinent contribution 
to photon radiation has been investigated previously in Ref.~\cite{Liu:2006imd} in the context of SPS direct-photon spectra at very low energies by the WA98 
collaboration~\cite{WA98:2003ukc}. In the calculations of Ref.~\cite{Liu:2006imd} the Bremsstrahlung's processes have been evaluated explicitly using $\sigma$ 
and $\rho$-meson pole-graphs for the elastic $\pi\pi$ scattering amplitude, and the rate was added incoherently as a separate emission source. Here, the in-
medium pion widths subsume these contributions  into the $\rho$ selfenergy (albeit more schematically from a microscopic point of view way); the magnitude 
and energy dependence of the enhancement in the pertinent photon rate due to the Bremsstrahlung contribution as shown in Fig.~\ref{fig_photon-rate} is very 
similar to what has been found Ref.~\cite{Liu:2006imd} (see, \eg, the right panel of Fig.~9 in there).

%%%%%%%%%%%%%%%%%%%%%%%%%%%%%%%
\section{Conductivity}
\label{sec_cond}
%%%%%%%%%%%%%%%%%%%%%%%%%%%%%%%
%
\begin{figure*}[!thb]
%\begin{minipage}[c]{0.32\linewidth}
%\vspace{-0.5cm}
%\hspace{-0.2cm}
%\includegraphics[width=1.0\textwidth]{sig_EM-lx}
%\end{minipage}
%\hspace{-0.2cm}
\begin{minipage}[c]{0.515\linewidth}
\vspace{-0.7cm}
\hspace{-0.0cm}
\includegraphics[width=1.0\textwidth]{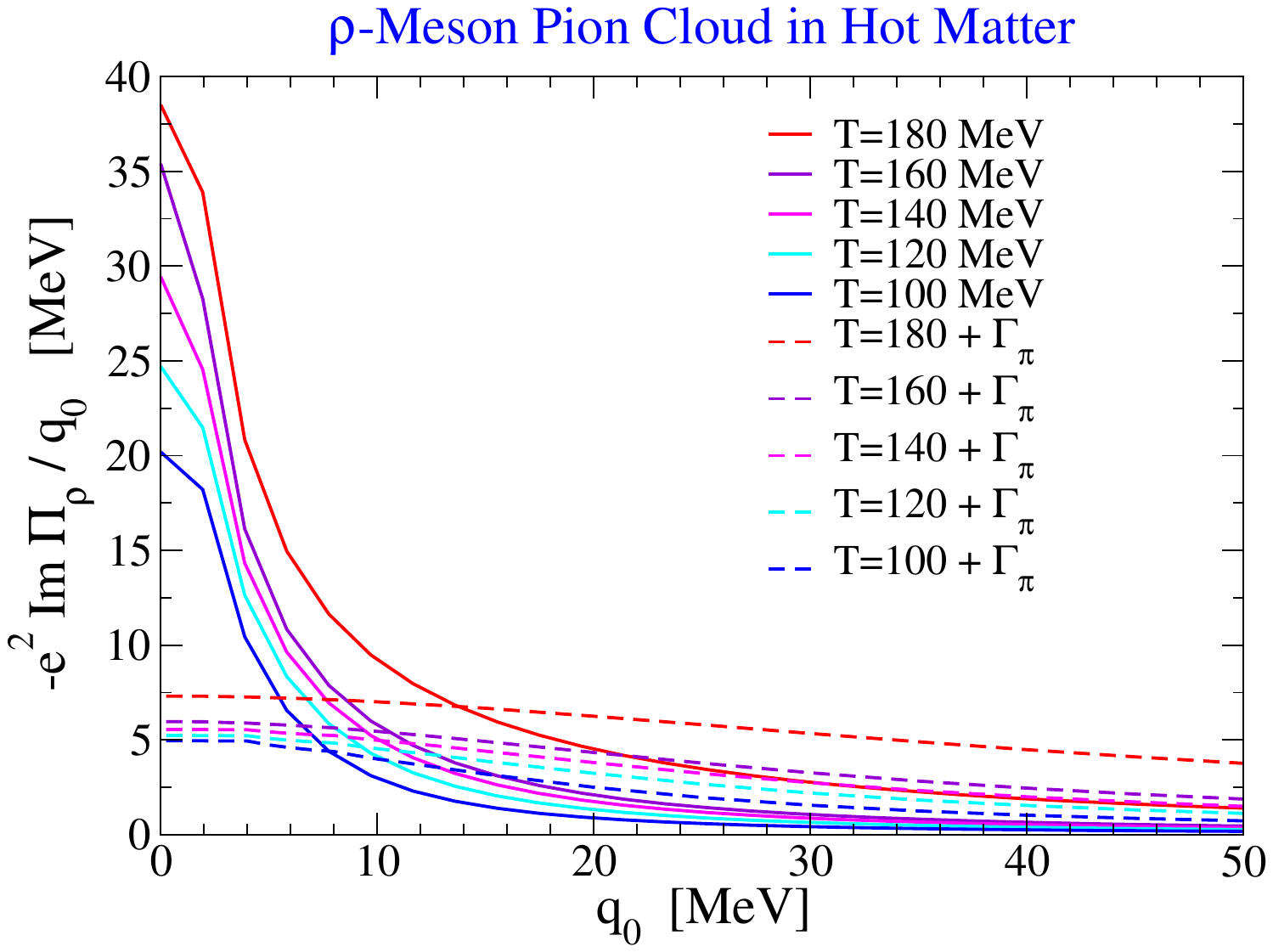}
\end{minipage}
\hspace{-0.8cm}
\begin{minipage}[c]{0.48\linewidth}
\vspace{-0.5cm}
\hspace{-0.2cm}
\includegraphics[width=1.0\textwidth]{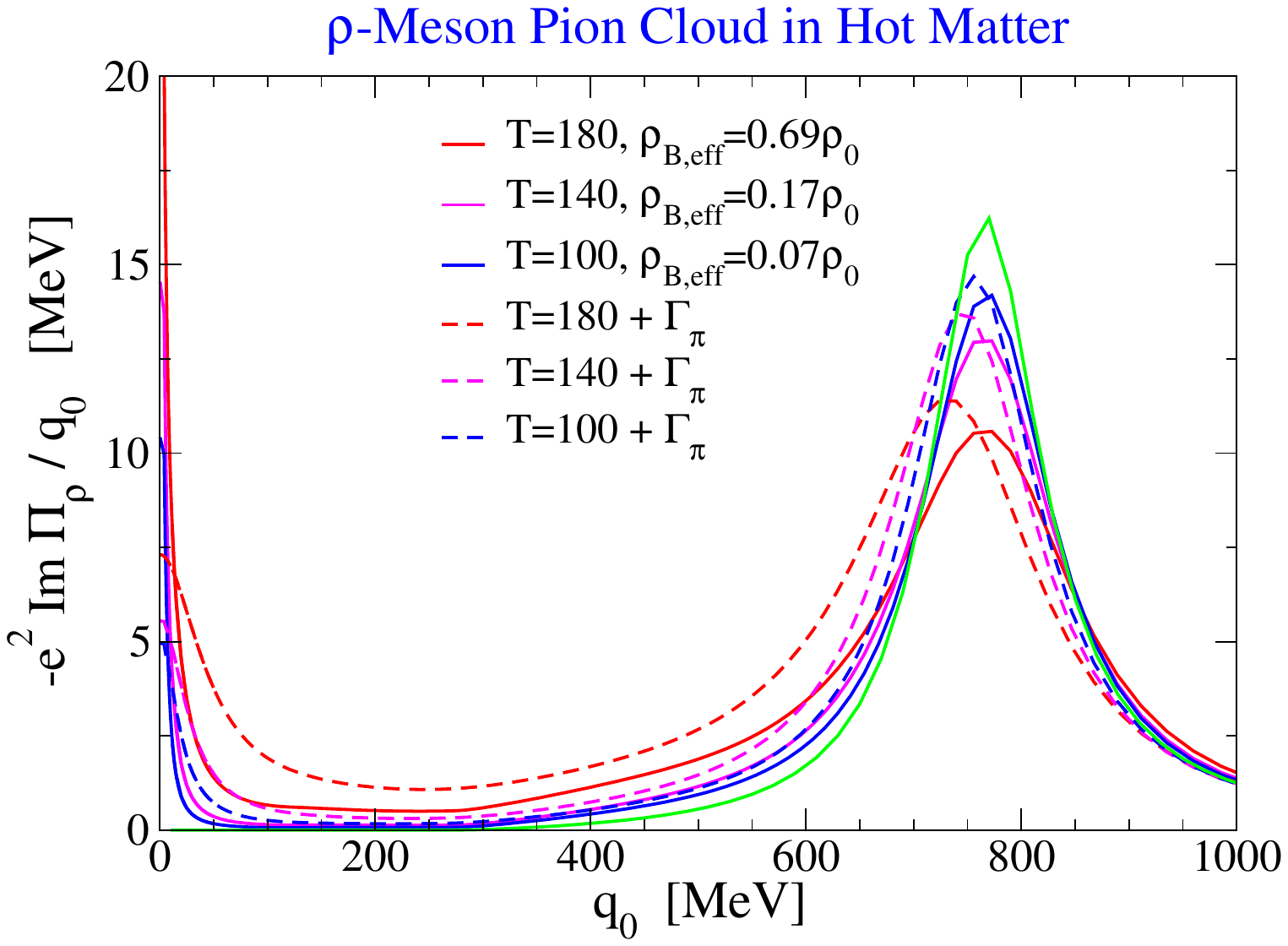}
\end{minipage}

\vspace{-0.3cm}

\caption{Hadronic EM spectral function in the isovector channel at total 3-momentum $q$=1\,MeV as obtained when only modifications of the 
$\rho$-meson's pion cloud are accounted for (left panel: very-low-mass region including the conductivity peak at vanishing energy); right panel: across 
the $\rho$-meson resonance region). Solid lines correspond to the full off-shell calculations when only including the medium modifications induced 
by anti-/baryons (plus Bose enhancement factors for the intermediate pions), while the solid lines additionally include thermal pion widths extracted 
from microscopic calculations using realistic $\pi\pi$ interactions.}
\label{fig_rhoem-picloud}
\end{figure*}
We now turn to the conductivity by exploiting its definition as the (timelike) zero-energy limit of the EM spectral function at vanishing 3-momentum, 
\be
\sig(T) = \frac{e^2}{2} \lim_{q_0\to 0} \rhoem(q_0,q=0)/ q_0  \  .
\ee
In the previous section we have already alluded to the importance of the pion cloud contributions to the low-energy part of the EM spectral function. 
It turns out that the direct $\rho$-meson interactions with mesons and baryons from the heat bath, which are implemented via resonance 
interactions (\eg,  $\rho+\pi \to a_1$ or $\rho+N\to N^*(1520)$, respectively), result in $\rho$-meson selfenergies that do not contribute to the 
conductivity. The reason for that is that the underlying interaction lagrangians for the three-point vertices need to satisfy current conservation which 
is achieved by using couplings to the $\rho$-meson field strength tensor, $\rho^{\mu\nu}$. The derivatives in this coupling result in selfenergies which are 
proportional to the square of the $\rho$'s energy, 3-momentum, or invariant mass~\cite{Rapp:1999ej}. In either case, it entails a vanishing contribution 
to the conductivity (this was also found for the charge-susceptibility, $\chi_{\rm ch}$, which can be extracted from the spacelike limit of the EM spectral 
function,  $\rhoem(q_0=0,q\to0)$~\cite{Prakash:2001xm}).   

We therefore focus on the pion cloud contributions to the $\rho$ selfenergy in this section. The low-energy 
region of the EM spectral functions, shown in Fig.~\ref{fig_rhoem-picloud} left, exhibits rather large values of $\sig\simeq20-40$\,MeV when only baryonic 
resonance excitations of the pions are included (currently limited to $P$-wave excitations $\pi+B_1\to B_2$). However, the inclusion of a constant pion 
width estimated from interactions with thermal pions markedly reduces the conductivity, by a factor of 4-5 down to values of $\sig\sim5-7.5$\,MeV, in line 
with the recent microscopic calculations in Ref.~\cite{Atchison:2022yxm}. At the same time the medium modifications in the $\rho$ resonance peak region 
due to  thermal pion widths are much less pronounced (see right panel of  Fig.~\ref{fig_rhoem-picloud}). Also recall that the total pion cloud modification, 
which is mostly due to the effects from baryons, makes up a $\sim$1/3 portion of the total broadening when also the direct $\rho$-hadron interactions 
are accounted for (see left panel of Fig.~\ref{fig_rhoem-x}). This suggests that the conductivity chiefly probes medium effects due to the light 
charge carriers in the hadronic medium, while the $\rho$ melting at higher masses is mostly driven by rescattering of the pions and $\rho$ mesons on 
the anti-/baryons in the medium.    
 
For the QGP spectral function, our calculation based on Ref.~\cite{Rapp:2013nxa} yields $\sig/T\simeq e^2 0.23 \Cem$,  
%at $T=1.45T_{\rm pc}\simeq 225$\,MeV (using $T_{\rm pc}\simeq 160$\,MeV 
which, using $C_{\rm EM}=6/9$ evaluates to 0.014, quite compatible with recent lQCD computations for temperatures $T\gsim 200$\,MeV, as reported in 
Refs.~\cite{Aarts:2014nba,Brandt:2015aqk,Ding:2016hua,Astrakhantsev:2019zkr} (note that in these works, the factor $e^2 \Cem$ is usually taken out, 
which for our results corresponds to 0.23), cf.~also Ref.~\cite{Aarts:2020dda}. 
The $\rho$ spectral functions based on Refs.~\cite{Urban:1999im,Rapp:1999us} lead to much larger conductivities, $\sig/T\simeq 0.2$, with a rather weak
temperature dependence. However,  when including the dressing of the $\rho$'s pion cloud via interactions with thermal pions, this value drops to 
$\sim$0.04, which is much closer to the lQCD result  (albeit still significantly larger).  However, in the vicinity of the pseudocritical region, lQCD 
computations lead to even smaller values,  $\sig/T \simeq \Cem e^2 0.05 \simeq 0.003$, with relative large systematic errors. At face value, these 
results are smaller than quantum lower bounds estimated from the AdS/CFT correspondence. The origin of the large discrepancy between the lQCD 
results and hadronic calculations in the pseudocritical region deserves further study. In the present context, we retain our more schematic model for 
the QGP emission where the difference between the QGP and hadronic results for is much smaller. Possible additional contributions in the hadronic 
calculations might arise from interactions with thermal kaons (which are typically at a level of 20\% of the pion contributions), $S$-wave pion-baryon 
interactions (which are usually much  smaller than the $P$-wave ones~\cite{Holt:2020mwf}), or gauge-invariant coupling schemes for direct 
$\rho$-hadron interactions that are not based on the $\rho$-meson field-strength tensor.

%%%%%%%%%%%%%%%%%%%%%%%%%%%%%%%
\section{Dilepton Spectra}
\label{sec_dl-spec}
%%%%%%%%%%%%%%%%%%%%%%%%%%%%%%%
%
\begin{figure*}[!thb]
\begin{minipage}[c]{0.32\linewidth}
\vspace{-0.5cm}
\hspace{-0.4cm}
\includegraphics[width=1.20\textwidth]{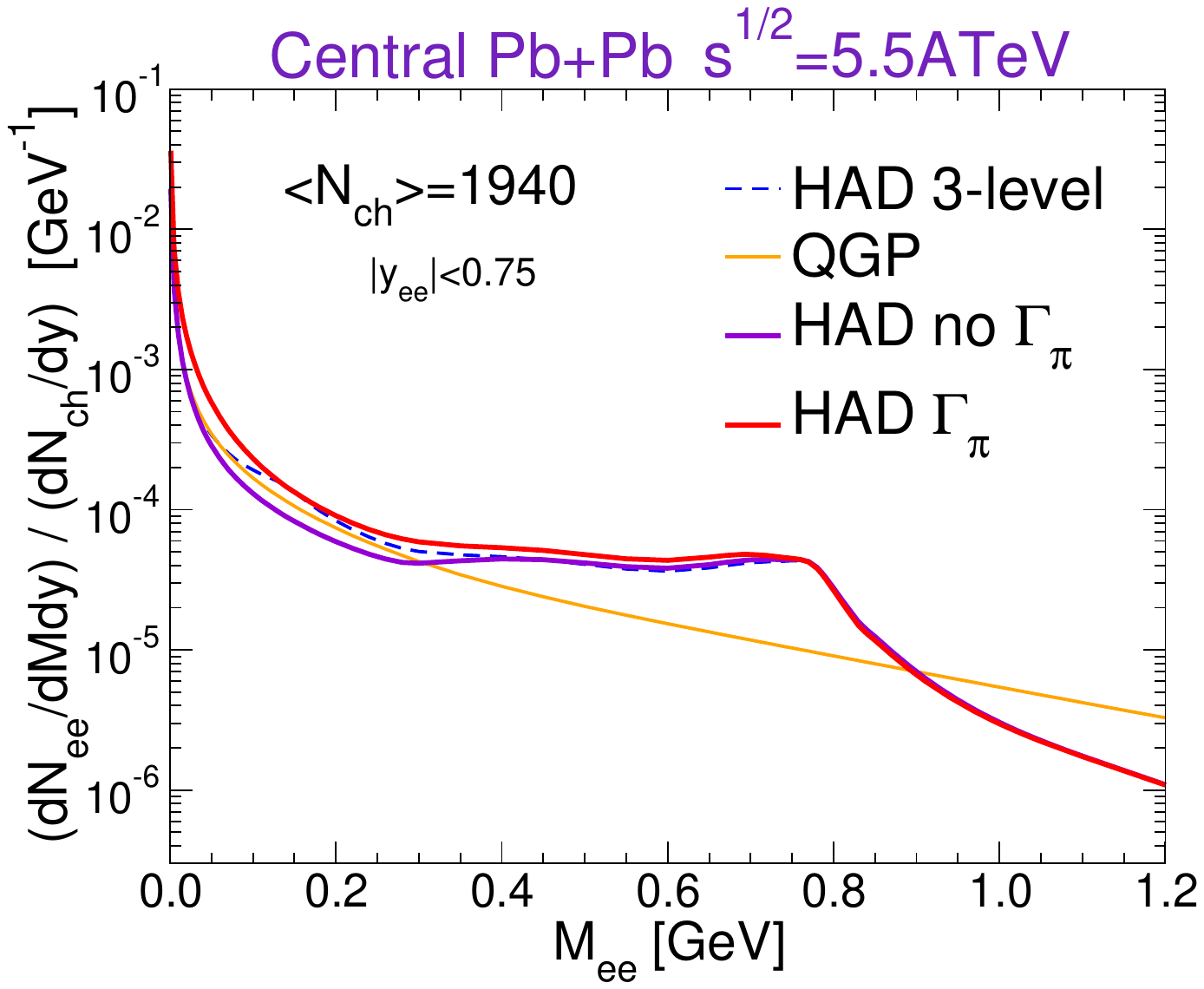}
\end{minipage}
\begin{minipage}[c]{0.32\linewidth}
\vspace{-0.5cm}
\hspace{-0.4cm}
\includegraphics[width=1.20\textwidth]{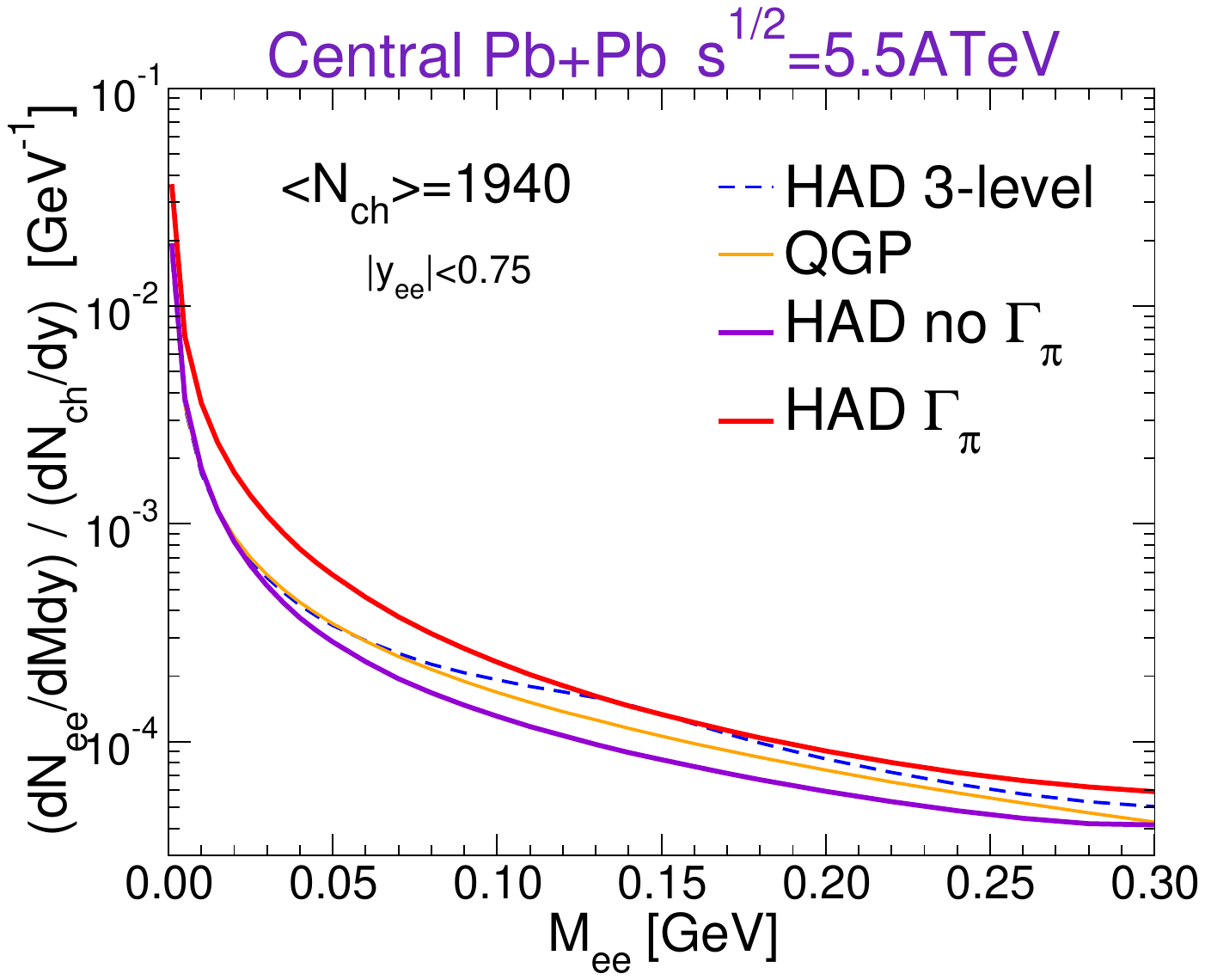}
\end{minipage}
\begin{minipage}[c]{0.32\linewidth}
\vspace{-0.5cm}
\hspace{-0.4cm}
\includegraphics[width=1.20\textwidth]{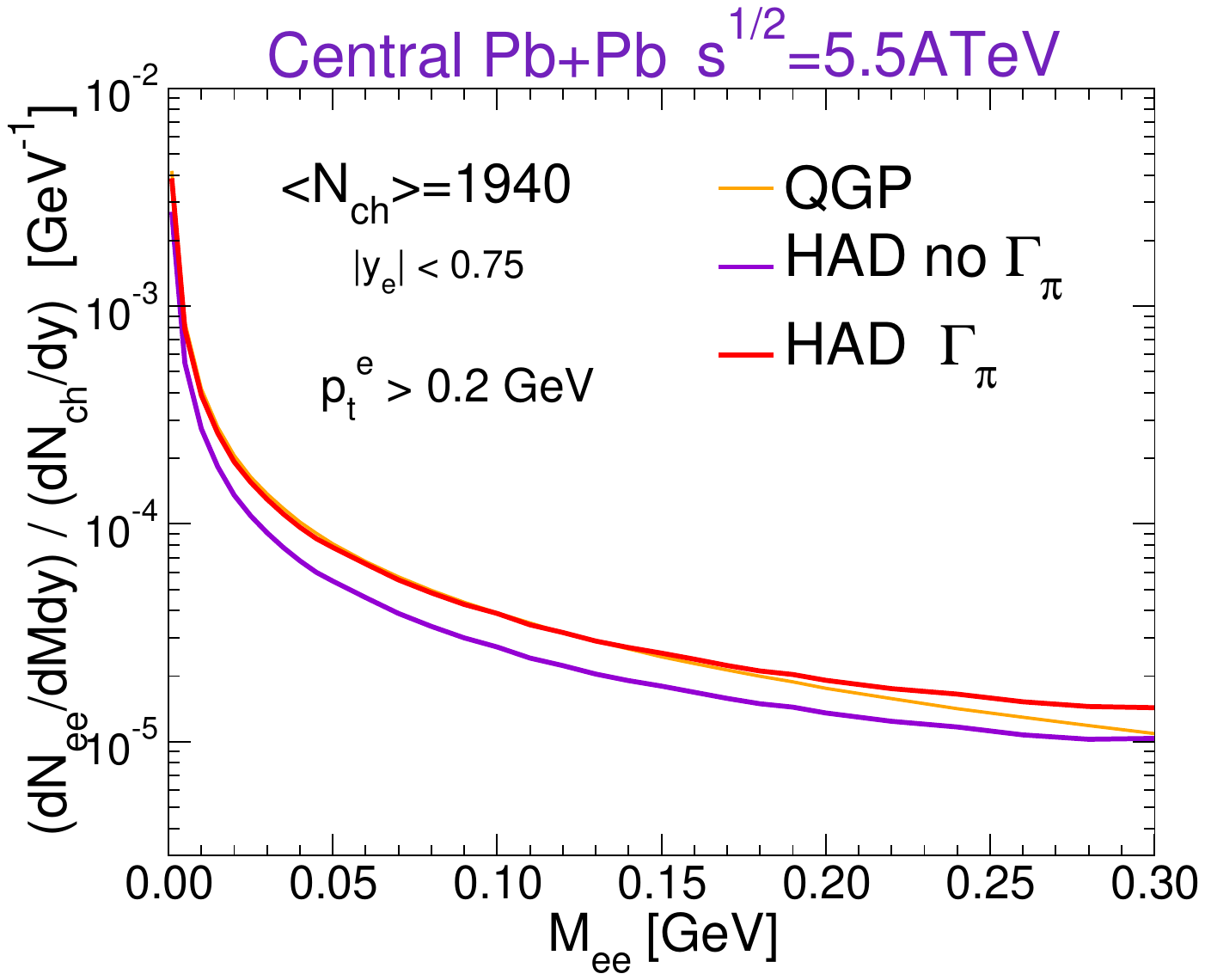}
\end{minipage}

\vspace{-0.3cm}

\caption{Three-momentum integrated low-mass dilepton spectra from thermal radiation in central Pb-Pb(5.02\,TeV) collisions, with only an acceptance cut on the pair rapidity applied; the left (middle) panel show the (very-) low-mass region, while the right  panel includes an extra cut on the single-electron $p_t$. The dotted line is the result for the previously used three-level approximation for the baryon-induced medium modifications in the pion cloud of the $\rho$, while the solid red and purple lines are using a full off-shell integration over the pion selfenergy with and without the addition of a pion width from interactions with 
pions in the heat bath, respectively; the orange line represents the QGP contribution.}
\label{fig_spec-0}
\end{figure*}
We now proceed to compute dilepton spectra in heavy-ion collisions at the LHC using thermal emission rates based on the spectral functions discussed 
in the preceding section (in addition to the in-medium $\rho$ we also include the isoscalar contribution from in-medium $\omega$ mesons and a 
multi-meson continuum at invariant masses above $\sim$1\,GeV; both contributions are, however, immaterial for the purpose of the present discussion). Following our earlier applications, we employ a schematic cylindrical fireball model assuming an isotropic medium with isentropic expansion over which we convolute the rates to obtain the spectra, 
\be
\frac{dN_{l^+l^-}}{dM} = \int \frac{Md^3q}{q_0} d^4x \Vfb(t) \frac{dN_{l^+l^-}}{d^4qd^4x} \ .
\ee
The expansion parameters (relativistic transverse and longitudinal accelerations and speeds) and total entropy, $S_{\rm tot}$, in the fireball volume, 
$\Vfb(t) $, are adjusted to reproduce blastwave results for the kinetic freezeout of light-hadron spectra as well as their experimentally observed 
multiplicities for a given centrality class, respectively. To convert the thus obtained time-dependent entropy density, $s(T) = S_{\rm tot}/\Vfb$ into 
a temperature requires the equation of state of the system. We utilize a parameterization of lQCD data above $T$=170\,MeV, smoothly matched to a 
hadron resonance gas below. Hadro-chemical freezeout is implemented at $(\mu_B,T_{\rm chem})\simeq (1,160)$\,MeV, after which effective chemical 
potentials are introduced for the light hadrons which are stable under strong decays  (including anti-baryons) to keep their ratios constant throughout 
the hadronic evolution.  Kinetic freezeout typically occurs around $T_{\rm kin}$=100\,MeV in central Pb-Pb collisions at the LHC.

Let us first inspect the 3-momentum integrated di-electron invariant-mass spectra (setting the electron mass to zero) with minimal restrictions on 
the measured kinematics of the lepton pair applied, \ie, merely a pair-rapidity cut around midrapidity of $|y_{ee}|<0.75$. From the left panel of 
Fig.~\ref{fig_spec-0} we see the usual feature that hadronic emission dominates in the $\rho$ resonance region with a strongly smeared-out peak, 
while QGP emission at the LHC takes over for invariant masses $M\gsim 1$\,GeV. At very-low masses, \ie, below the 2-pion threshold, $M\lsim 300$\,MeV, 
both contributions are comparable. Finer details can be seen when zooming in to this region, cf.~middle panel of Fig.~\ref{fig_spec-0}. 
Within the previously employed 3-level approximation for the baryon-induced selfenergies of the pions, hadronic emission is slightly larger, by up 
to $\sim$20\% or so,  around $M\simeq150$\,MeV (note that above the two-pion threshold there is little difference between the full calculation and the 
3-level approximation). However, with the more accurate full off-shell calculations~\cite{Urban:1999im}, the hadronic emission is mostly below the 
QGP emission by up to 20\% or so.  Finally, when including additional pion widths estimated from interactions with the surrounding pion gas, 
we are back to a situation where hadronic emission is slightly stronger, with the enhancement now growing toward smaller masses, amounting, \eg, to 
$\sim$50\% at $M$=50\,MeV.  Recalling that the conductivity is much smaller for the case with thermally induced pion widths, one can conclude that 
dileptons at very low masses are indeed quite sensitive to the electric conductivity: a factor of about 5 {\em reduction} in the conductivity 
leads to an {\em enhancement} of more than a factor of 2 in the 3-momentum integrated dilepton spectrum at $M$=50\,MeV (for the hadronic contribution). 
This implies that, much like for the photon rates, the observable effect of changes in the conductivity is the broadening of the conductivity peak, not its 
reduction at the zero-energy intercept. However, when applying a rather common experimental acceptance cut where the transverse momentum of the 
individual electron and positron tracks is restricted to $p_t^e>0.2$\,GeV, the sensitivity to the conductivity reduces appreciably, \ie, the enhancement at $M$=50\,MeV is only at a level of $\sim$50\%.

\begin{figure*}[!thb]
\begin{minipage}[c]{0.32\linewidth}
\vspace{-0.5cm}
\hspace{-0.4cm}
\includegraphics[width=1.2\textwidth]{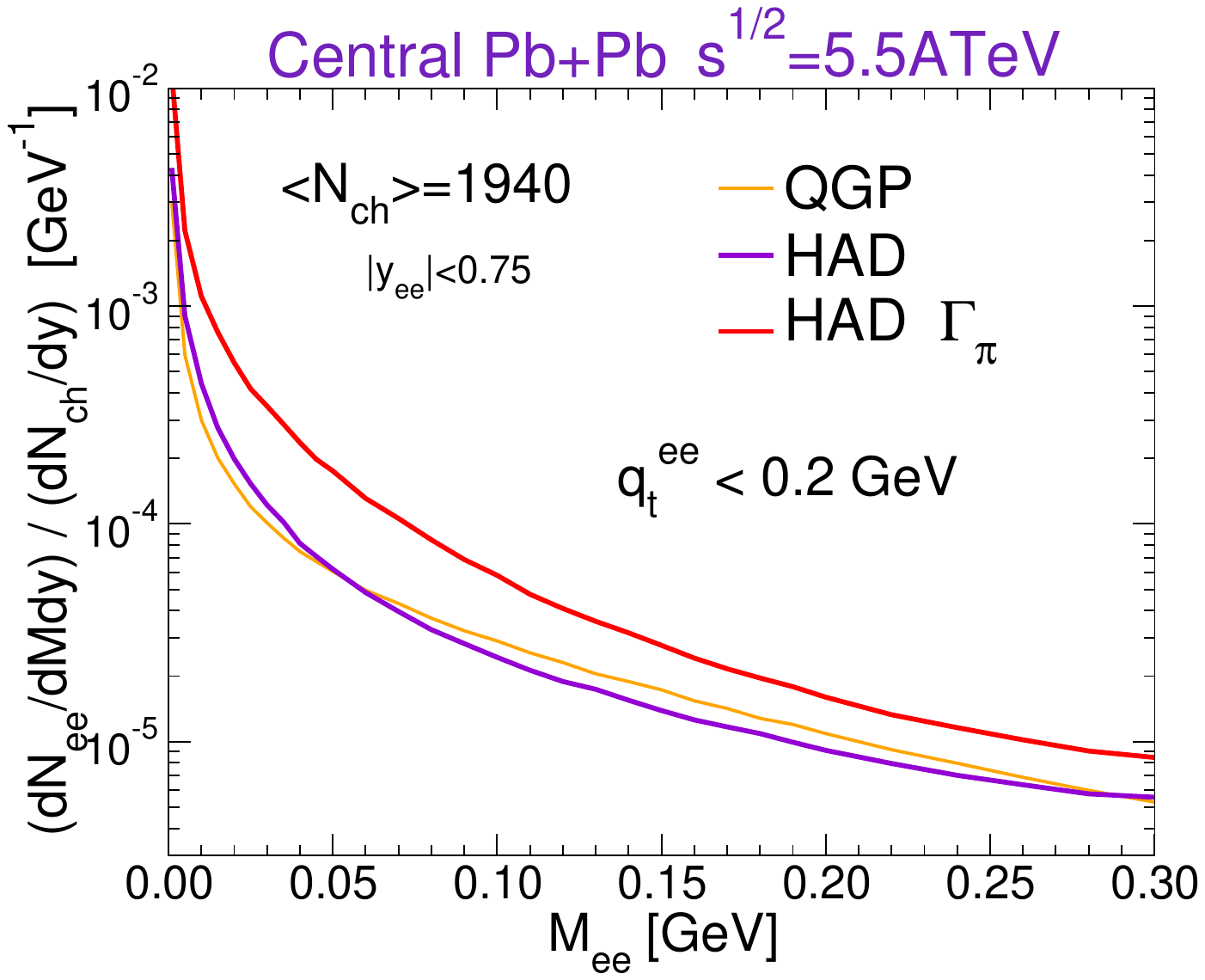}
\end{minipage}
\begin{minipage}[c]{0.32\linewidth}
\vspace{-0.5cm}
\hspace{-0.4cm}
\includegraphics[width=1.2\textwidth]{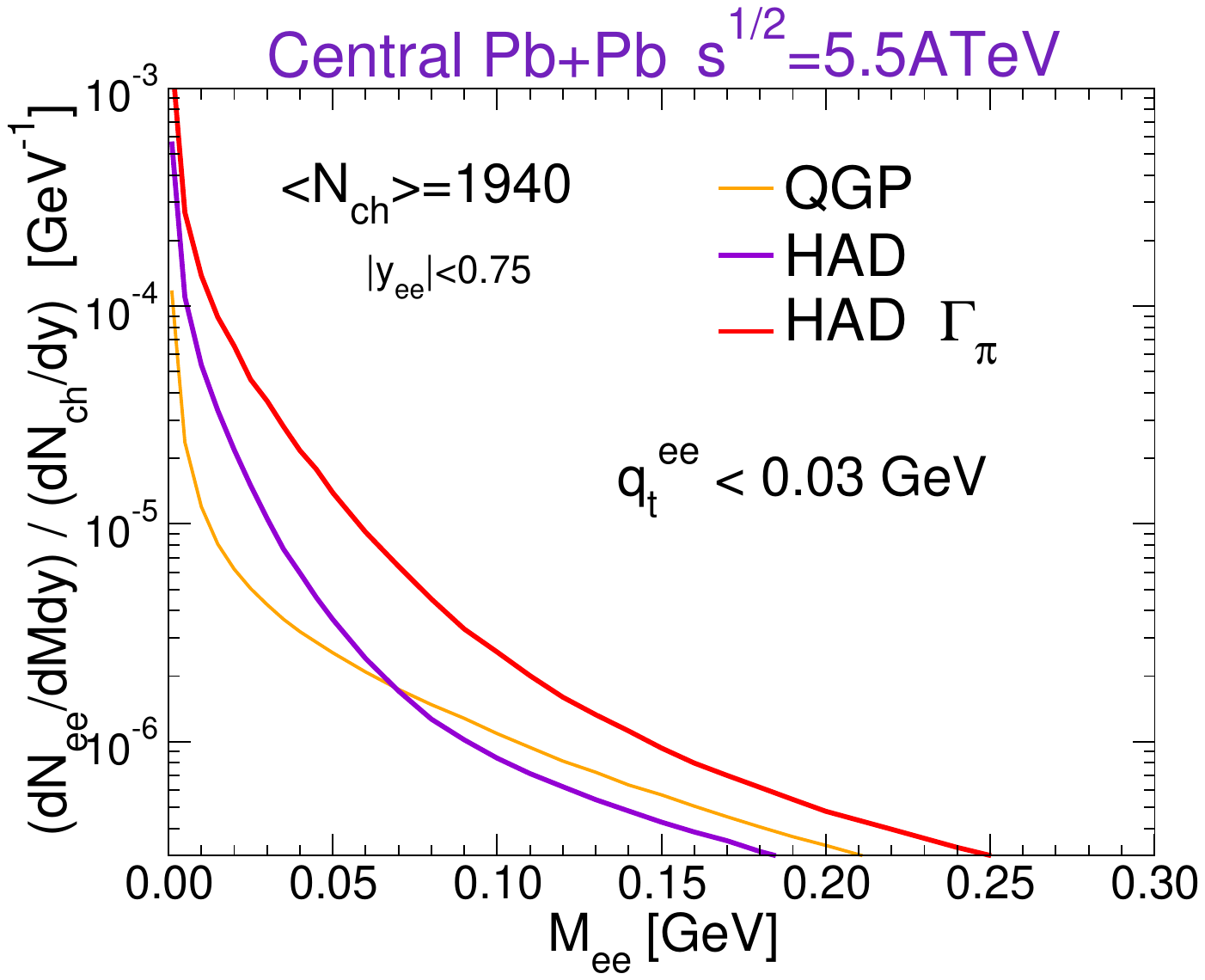}
\end{minipage}
\begin{minipage}[c]{0.32\linewidth}
\vspace{-0.5cm}
\hspace{-0.4cm}
\includegraphics[width=1.2\textwidth]{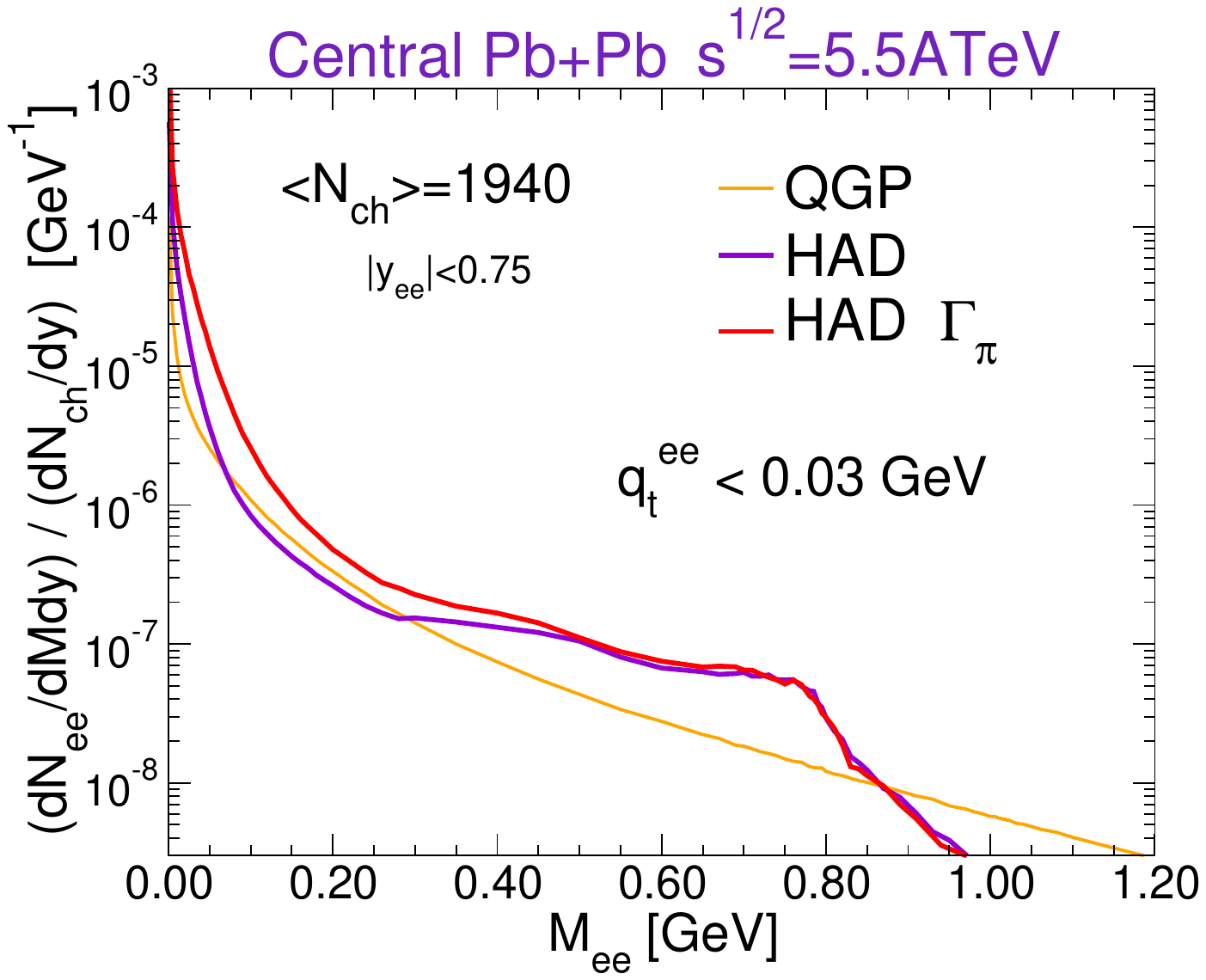}
\end{minipage}

\vspace{-0.3cm}

\caption{Very low-mass dilepton spectra in central Pb-Pb collisions at the LHC when applying an upper limit on the pair momentum of $q_t^{ee}<0.2$\,GeV 
(left panel) and 0.03\,GeV (middle panel); note the reduction in the overall yield for the latter case, which is also shown over a wider mass region in the right panel, relative to the left panel in Fig.~\ref{fig_spec-0}. }
\label{fig_spec-cuts}
\end{figure*}
Recalling the EM spectral functions in Fig.~\ref{fig_rhoem-x}, which indicate that the conductivity peak has essentially disappeared for a 3-momentum of 
$q$=300\,MeV (note that at this momentum,  $M$=0 corresponds to a photon energy of 300\,MeV), we investigate the impact of an upper cut on the 
electron pair's (or virtual photon's) transverse momentum, $q_t^{ee}$, in Fig.~\ref{fig_spec-cuts}.  Note that this differs from the 3-momentum, $q$, 
of the EM spectral function in the thermal restframe by a Lorentz-boost caused by the collective expansion of the fireball medium. Nevertheless, when 
$q_t^{ee} < 0.2$\,GeV is imposed,  the enhancement from the scenario with smaller conductivity due to the thermal pion widths increases the 
hadronic dilepton yield at $M$=50\,MeV to a factor of 3 (compared to $\sim$2 without the cut), see left panel in Fig. ~\ref{fig_spec-cuts}. With a more 
extreme choice of  $q_t^{ee} < 0.03$\,GeV, the enhancement is close to a factor of 4. There is, of course a price to pay in that the pertinent yields
in the spectra are substantiality suppressed, by a factor of  $\sim$3 for $q_t^{ee} < 0.2$\,GeV and about an additional order of magnitude for 
$q_t^{ee} < 0.03$\,GeV. An optimized experimental sensitivity will thus have to balance an enhancement due to the reduced conductivity with the 
loss of total yield.

%%%%%%%%%%%%%%%%%%%%%%%%%%%%%%%%%%%%%%%%%%%%%%%%%%%%%%%%%%%%%%%%%%%%%%%
\section{Conclusions}
\label{sec_concl}
%%%%%%%%%%%%%%%%%%%%%%%%%%%%%%%%%%%%%%%%%%%%%%%%%%%%%%%%%%%%%%%%%%%%%%%
In this paper we have studied the electric conductivity of hot QCD matter and its manifestation in thermal dilepton spectra in ultarelativistic heavy-ion 
collisions, utilizing their formal connection through the EM spectral function. In the hadronic phase, we have employed in-medium vector-meson spectral 
functions computed in hadronic many-body theory in connection with the vector dominance model to couple to the EM current; in particular, we have 
improved the calculation of the in-medium pion cloud of the $\rho$-meson's selfenergy through an explicit off-shell treatment which turned out to be 
quantitatively important for the very-low mass region of the EM spectral function. Furthermore, the addition of thermal pion widths was found to produce 
the most relevant contribution to broadening the conductivity peak near zero energy. This, in particular, suggests that the light degrees of freedom in hadronic 
matter are the key carriers of the EM current, while in the $\rho$ resonance region baryon-induced effects are dominant. For the quark-gluon plasma, 
we have used a perturbative $q\bar q$ spectrum augmented with a conductivity peak that matches lattice QCD data at temperatures above $\sim200$\,MeV. 
At lower temperatures the lattice data show a rather strong decrease in $\sig/T$ that leads to a marked discrepancy with our current hadronic calculations 
which requires further study (\eg, by scrutinizing the in-medium selfenergies of the pions).
%, and has not been implemented into our QGP model (yet). 
When applying this setup to compute thermal dilepton spectra we have found that the chief signature of a small conductivity is an enhancement in the 
dilepton yields in the mass region below the 2-pion threshold, caused by the broadening of the conductivity peak at zero energy. Since the latter rapidly
ceases at finite 3-momentum, the signal can be enhanced by applying upper cuts on the lepton pair momentum, while in practice this will have to be 
balanced by the accompanying strong reduction in the total yields. Overall, the sensitivity of the very-low-mass region in thermal dilepton radiation is
promising for a meaningful measurement with future precision data at ALICE-3 in run-5 at the LHC. In the meantime, efforts are 
underway to improve the theoretical knowledge about the magnitude of the conductivity in the pseudocritical region.

%%%%%%%%%%%%%%%%%%%%%%%%%%%%%%%%%%%%%%%%%%%%%%%%%%%%%%%%%%%%%%%%%%%%%%%
{\it Acknowledgments.--}
 %\label{sec_acknow}
%%%%%%%%%%%%%%%%%%%%%%%%%%%%%%%%%%%%%%%%%%%%%%%%%%%%%%%%%%%%%%%%%%%%%%%
This work has been supported by U.S.~National Science Foundation through grant no.~PHY-2209335.  
The author gratefully acknowledges Michael Urban for providing his code for the off-shell calculations of the in-medium pion cloud of the $\rho$ meson.

\bibliography{ref-em}

\end{document}